\documentstyle[prl,aps]{revtex}
\input epsf
\def\curl{\mathop{\rm curl}\nolimits}

\begin{document}
\draft
\twocolumn[
\widetext
\title{Effect of the vortex core on the magnetic field in hard
       superconductors}
\author{A. Yaouanc and P. Dalmas de R\'eotier}
\address{Commissariat \`a l'Energie Atomique,
\\ D\'epartement de
   Recherche Fondamentale sur la Mati\`ere Condens\'ee,
\\ Service de Physique Statistique, Magn\'etisme et Supraconductivit\'e,
   F-38054 Grenoble cedex 9, France}
\author{E.H. Brandt}
\address{Max-Planck-Institut f\"ur Metallforschung,
         D-70506 Stuttgart, Germany}
\date{\today} \maketitle \widetext
\leftskip 54.8pt
\rightskip 54.8pt
\begin{abstract}
Using approximate analytical and new numerical solutions of the
conventional Ginzburg-Landau equations we calculate the small angle
neutron scattering cross-section and the variance of the field
distribution as measured by muon-spin rotation for superconductors
with large Ginzburg-Landau parameter $\kappa$.
Our results prove that a proper account of the
finite size of the vortex core is important, even at relatively low
fields. This finding provides a natural explanation for the recently
observed field dependence of the CeRu$_2$ form factor and
of the YBa$_2$Cu$_3$O$_{6.95}$ penetration depth.
\par
\end{abstract}
]
\narrowtext

The study of the vortex state in high temperature and heavy fermion
superconductors is presently a subject of intense investigation.
Numerous publications are devoted to the measurement of the magnetic
penetration length $\lambda$ since this is one way to probe the nature
of the low energy excitations and the symmetry of the pairing state.
Among the possible experimental techniques available to investigate
the vortex lattice, small angle neutron scattering (SANS) and muon spin
rotation ($\mu$SR) experiments are unique since they directly probe the
bulk of the material and allow to determine not only the
field and temperature dependence of $\lambda$ but also its value
at low temperature, see the recent
Refs.\ \cite{Keimer94,Sonier,Kleiman,Broholm90,Yaouancupt3,Huxley95}.
To extract quantitative information from SANS and $\mu$SR
measurements, a detailed theory of the magnetic field
inside the superconductor is needed, going beyond the London model
which treats the vortex cores as mathematical singularities.
The finite core size was considered in Refs.\ \cite{Clem,Hueb,Hao}.
\par
In this paper we compute the Fourier components of the magnetic
field in a type-II superconductor containing an ideal vortex lattice.
We disregard pinning \cite{Brandtpinning} and vortex ``phases''
such as the glassy or liquid states \cite{Blatter94,Brandt95}.
When accounting for the finite size of the vortex cores within
the Ginzburg-Landau (GL) theory we find an unexpected large reduction
of all Fourier components down to very low inductions $B$. Although
our results are based on the conventional GL theory, they still are
of relevance for the analysis of unconventional superconductors
such as high $T_c$ superconductors and heavy fermion superconductors.
For example, in recent reports \cite{Joyntupt3,Affleck}
the effect of the finite size of the vortex core is described as
if these compounds were conventional superconductors.
\par
We define an orthogonal reference frame ($x$,$y$,$z$), with the
external magnetic field ${\bf B}_{\rm ext}$ applied along the $z$ axis
chosen along one of the three main axes  $\bf a$, $\bf b$ and $\bf c$
of the penetration-length tensor such that the vortices are also
along $z$. For superconductors with large
GL parameter $\kappa = \lambda/\xi \gg 1$ ($\xi$ is the coherence
length) at not too large fields $B_{\rm ext} \ll B_{c2}$
($B_{c2}$ is the upper critical field) we may approximate the
vortex fields by the London model. The London field ${\bf B(r)}$
caused by straight vortices located at
sites ${\bf r}_v$ satisfies \cite{Balatskii86,Schopohl88}
\begin{eqnarray}     
{\bf B(r)} + \curl [\Lambda \curl {\bf B(r)}] & = &
\Phi_0 \sum_v \delta( {\bf r -r}_v)\, {\bf \hat z} \,.
\label{field1}
\end{eqnarray}
Here  $\Phi_0$ = 2.07 $\times$ 10$^{-15}$ Tm$^2$ is the quantum
of flux, the sum is over the vortices,  $\delta({\bf r})$ is
the two-dimensional delta function, and ${\bf \hat z}$ is the
unit vector along the vortex cores. The eigenvalues of the tensor
$\Lambda$ are expressed in terms of penetration lengths:
$\Lambda_a$ = $\lambda_a^2$, $\Lambda_b$ = $\lambda_b^2$, and
$\Lambda_c$ = $\lambda_c^2$. Here $\lambda_a$, $\lambda_b$ and
$\lambda_c$ are the penetration lengths for
currents flowing along the $a$, $b$ and $c$ axes, respectively.
\par
When the vortices form a regular lattice it is convenient to
introduce the Fourier components ${\bf B(G)}$ =
 $ \int {\bf B(r)} \exp(-i{\bf G \cdot r}) d^2 {\bf r}/S$
of the periodic magnetic field
 ${\bf B(r)}$ = $\sum_{\bf G} {\bf B(G)} \exp(i{\bf G \cdot r})$,
where $\bf G$ are the vectors of the reciprocal lattice and $S$ the
surface of the vortex lattice unit cell.
The London equation is then easily solved for the cases of main
interest, namely, ${\bf B}_{\text{ext}}$ parallel to either
 $\bf a$, $\bf b$ or $\bf c$.\  For these three geometries one finds
  \begin{eqnarray}   
  B_z( {\bf G}) & = & {\Phi_0 \over S}\, {1\over 1 +\Lambda_x G^2_y
  +\Lambda_y G^2_x}\,,
  \label{field6}
  \end{eqnarray}
and $B_x( {\bf G})$ = $B_y( {\bf G})$ = 0. Therefore,
as expected, there is no transverse field component. Equation
(\ref{field6}) means, for example, that if
${\bf B}_{\text{ext}}$ $\parallel$ $\bf c$ we write this equation
with $x = a$, $y = b$, and $z = c$. In this way we recover the result
of Ref. \onlinecite{Barford} for a uniaxial superconductor, in which
two penetration lengths are equal.
\par
Equation (\ref{field1}) disregards the effect of the finite size of
the vortex core, which removes the logarithmic infinity of
$B_z({\bf r})$ at ${\bf r}_v$ and thus reduces the amplitude of
the higher Fourier components. At $B \ll B_{c2}$ this effect is
accounted for by multiplication of the London
solution (\ref{field6}) by a cutoff factor. Here a general remark
seems appropriate. There is no general theory of $\bf B(r)$ valid at
arbitrary temperature, and even if it existed (if
the BCS-Gorkov-Eliashberg theory would apply and could be solved)
the material parameters entering such a theory are not known with
sufficient accuracy, {\it e.g.\/}, the anisotropic electron mean free
path $l$, the shape of the Fermi surface, and the coupling constant.
Even when we use the Ginzburg-Landau (GL) theory to obtain a cutoff,
we find that a general analytical solution does not exist, not even
in the limit $\kappa \to \infty$, which would be sufficient here.
If the GL theory is applicable it applies down to $B=0$.
Below we derive the low-field cutoff factor from approximate
analytical solutions of the GL theory and from a numerical solution.
We are considering first an isotropic superconductor.
\par
The best analytical GL expression available
was obtained by Clem \cite{Clem} for isotropic superconductors
at low inductions $B\ll B_{c2}$. Using a Lorentzian trial function
for the order parameter $|\psi({\bf r})|^2 $ of an isolated vortex,
Clem finds for large $\kappa \gg 1$
  \begin{eqnarray}    
  B_z(G) = {\Phi_0 \over S} {g\, K_1(g) \over 1+\lambda^2 G^2},
   ~~~ g = \sqrt 2\, \xi ( G^2 + \lambda^{-2} )^{1/2} \,.
  \label{clem}
  \end{eqnarray}
Here $K_1(x)= -K_0'(x)$ is a modified Bessel function with the
limits $K_1(x) = 1/x -(x/2)\ln (1.7139/x)$ ($x\ll 1$)
and    $K_1(x) = (\pi/2x)^{1/2} \exp(-x)$  ($x\gg 1$). From Eq. (\ref{clem})
we recover the London solution if the cores size shrinks to zero.
The cutoff factor $gK_1(g)$ in Eq. (\ref{clem}) may be approximated for all
$g$ values by $\exp(-\sqrt 2 \xi G)$, or, less accurate but
convenient for computations, by $\exp( -2\,\xi^2 G^2)$ as
suggested in Ref. \onlinecite{Brandt92}. The cutoff
$\exp(-\xi^2 G^2 /4)$ given in Ref. \onlinecite{Brandt72}
was derived from the GL solution near $B_{c2}$,
and is not valid at
low $B$ ($B\ll B_{c2}$). At intermediate fields the cutoff should
interpolate between these two expressions.
Therefore, the argument of the Gaussian cutoff used recently
is smaller than the one we propose: 1/4 \cite{Joyntupt3} or 1/2 \cite{Affleck}
instead of 2 valid at low $B$. The correct low-field cutoff
yields a stronger field dependence of the SANS intensity than
predicted for example in Ref. \onlinecite{Joyntupt3}.
\par
   Clem's approximate analytical theory of the dilute vortex lattice
was extended to larger fields and to anisotropic superconductors
by Hao {\it et al.} \cite{Hao}  using the same type of variational approach.
The resulting Fourier components for an isotropic superconductor
may be written as
  \begin{eqnarray}    
  B_z({\bf G}) & = & {\Phi_0 \over S}
{f_{\infty} K_1 \left[ {\displaystyle{ \xi_v \over \lambda}}
( f_{\infty}^2 + \lambda^2 G^2)^{1/2} \right]
\over ( f_{\infty}^2 + \lambda^2 G^2)^{1/2}
K_1 \Big( {\displaystyle{ \xi_v \over \lambda}} f_{\infty} \Big)},
  \label{cutoff1}
  \end{eqnarray}
where $\xi_v$ and $f_{\infty}$ are two variational parameters
representing the effective core radius of a vortex and the
depression of the order parameter due to the overlap of vortex cores,
 respectively. For the cases of interest here ($\kappa$ $>$ 10)
the two variational parameters have simple functional dependences
on $b$ $\equiv$ $B/B_{c2}$ and $\kappa$ \cite{Hao,Clem} :
  \begin{mathletters} 
  \label{cutoff2}
  \begin{eqnarray}
f_{\infty}^2 & = & 1 - b^4,
  \end{eqnarray}
  \begin{eqnarray}   
\xi_v  & = & \xi \Big( {\sqrt 2} - {0.75 \over \kappa} \Big)
 ( 1 + b^4 )^{1/2}  \left[ 1- 2 b (1-b)^2 \right]^{1/2}\!.
  \end{eqnarray}
  \label{cutoff3}
  \end{mathletters}
\noindent In Eqs.\ (\ref{cutoff2}) $\Phi_0/S=B=b B_{c2}$ is the mean
induction, which for
$2b\kappa^2 > 1$ may be equated to $B_{\rm ext}$.\par

For $\kappa \gg1$ the argument of $K_1$ in the {\it denominator}
of Eq.\ (\ref{cutoff1}) is much smaller than 1, thus we may use
$K_1(x) \approx 1/x$. Since for high $T_c$
superconductors and typical $B_{\rm ext}$ values,
$b$ is never larger than a few \%, we may also neglect the field
dependence of $f_{\infty}$ and $\xi_v$, putting
$f_{\infty}$ $\approx$ 1 and $\xi_v$ $\approx$ $\sqrt 2 \xi$. For
the analysis of measurements performed on heavy fermion
superconductors, the field dependence of $f_{\infty}(b)$ can thus
be disregarded (usually $B_{\rm ext}$ $\leq$ 1 T \cite{Kleiman})
but this may not be true for $\xi_v(b)$. For example, with UPt$_3$
at $B_{\rm ext}$ = 1 T one has $b$ $\approx 0.4$ \cite{Kleiman} and
therefore $\xi_v$ $\approx$ $0.854 \times  \sqrt 2 \xi$.
\par
\begin{figure}  
\centerline{\epsfbox{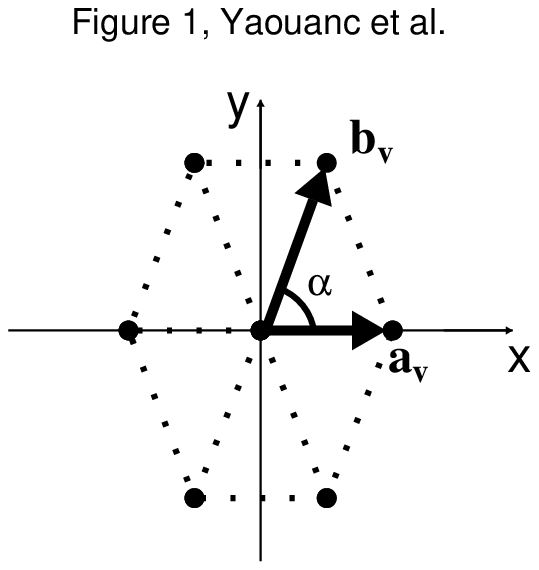}}
\caption[]{Definition of the primitive cell vectors ${\bf a}_v$ and
${\bf b}_v$ and angle $\alpha$ of a distorted vortex lattice in
real space. $\alpha$ is $\pi/2$ minus the angle defined in
Fig.\ 2 of Ref. \protect \onlinecite{Kleiman}.}
\label{lattice}
\end{figure}

The smallest non-zero reciprocal vector for an equilateral triangular
lattice is $G_{10}$ = $G_{\text{min}}$ = $a^*_{v}$ = $(2\pi /S) a_{v}$
(see Fig. \ref{lattice} for the definition of $a_{v}$), thus
$G^2_{\text{min}}$ = $(8 \pi ^2 / \sqrt{3}) (B/ \Phi_0)$.
This means that for the high $T_c$ compounds at $B$ $\simeq$
$B_{\text{ext}}$ = 20 mT one has
$\Lambda G^2_{\text{min}}$ $\simeq$ 10 $\gg$ 1, if
$ \Lambda^{1/2}$ = $\lambda = 1500$ \AA\ is used. For UPt$_3$
$\lambda$ is even larger \cite{Kleiman}. Accounting for the large
value of $\Lambda G_{\text{min}}^2=(4\pi/\sqrt3)b\kappa^2$ we may write
  \begin{eqnarray}   
  B_z({\bf G}) & = & {\Phi_0 \over S} {f_\infty^2 \over \Lambda G^2}
            \, (\xi_v G) \,  K_1 \left(\xi_v G \right).
  \label{cutoff4}
  \end{eqnarray}
In this letter we test the applicability of formula (\ref{cutoff4})
to recently published SANS results on CeRu$_2$.
\par
\begin{figure}  
\epsfbox{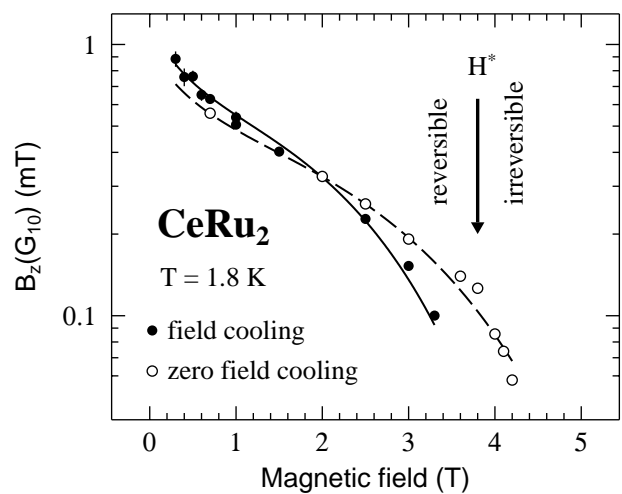}
\caption[]{Form factor for the reflection [1,0] from the vortex
lattice of CeRu$_2$ as a function of the applied field. The points
taken from Ref. \protect \onlinecite{Huxley95} have been obtained
using either a field cooling or zero field cooling procedure.
The lines are fits to Eq. (\protect \ref{neutron1}).}
\label{CeRu2}
\end{figure}

The conventional superconductor CeRu$_2$ has attracted some interest because
of its complex phase diagram in the $\left( B_{\text{ext}},T \right)$ plane.
Notably, a reversible-irreversible line is observed. The form factor $B_z(G)$
is easily obtained from the SANS cross-section \cite{deGennes}.
The CeRu$_2$ measurements of $B_z(G_{10})$ as a function
of $B_{\text{ext}}$ are presented in Fig. \ref{CeRu2}. Because $\Lambda$
is scalar, we derive from Eq. (\ref{cutoff4})
  \begin{eqnarray}  
  B_z(G_{10}) & = & {3^{1/4} \over 2 \pi \sqrt{2}}
{\sqrt {\Phi_0 B_{\text{ext}} } f_{\infty}^2  \xi_v \over \lambda^2} \cr
& \times &
K_1 \Big( {2 \pi \sqrt{2} \over 3^{1/4}} \xi_v \sqrt{ B_{\text{ext}} /
\Phi_0} \Big).
  \label{neutron1}
  \end{eqnarray}
This expression depends only on the two parameters $\lambda$ and
$\xi$. The fits yield for the data recorded either in field cooling
(FC) or zero field cooling (ZFC) procedure,
$\lambda$ = 1870 \AA\ and $\xi$ = 84 \AA\ and
$\lambda$ = 2090 \AA\ and $\xi$ = 74 \AA, respectively.
Taking the traditional point of view,
 the FC data reflect the equilibrium properties of the vortex lattice.
 From these data $\kappa = 22$ is larger than the previously estimated
$\kappa$ = 14.5 \cite{Huxley93}, \cite{Extra}. From the $\xi$ value we
compute $B_{c2}$ = $\Phi_0/ (2 \pi \xi^2)$ = 4.7 T. Magnetization
measurements at 1.8 K give
$B_{c2}$ = 5.3 T \cite{Huxley93}. The values deduced from the FC
neutron data are satisfactory in view of the well known difficulty
to extract a reliable $\kappa$ value from magnetization measurements.
\par
The traditional Gaussian cutoff predicts $\ln [B_z(G_{10})]$
$\propto B_{\text{ext}}$, {\it i.e.\/} a straight line in
Fig. \ref{CeRu2}. This is not observed.
\par
The generalization of Eq.~(\ref{cutoff4}) to anisotropic penetration
length tensors reads for $\kappa$ $\gg$ 1
  \begin{mathletters} 
  \label{cutoff6}
  \begin{eqnarray}
  B_z({\bf G}) & = & {\Phi_0 \over S}
{\left( 1 -b^4 \right)}
{u \cdot K_1(u) \over \Lambda_x G_y^2 + \Lambda_y G_x^2} \,.
  \end{eqnarray}
Here $uK_1(u)$ is an anisotropic cutoff factor with
  \begin{eqnarray}   
 u^2 & = & 2 \left( \xi_x^2 G_x^2\! +\! \xi_y^2 G_y^2 \right)
 \left( 1\! +\! b^4 \right)
 \left[ 1- 2 b \left(1\!-\!b \right)^2 \right],
  \end{eqnarray}
  \begin{eqnarray}   
 uK_1(u) \approx 1 -(u^2/4) \ln (2.937 / u^2) ~~~
                                {\rm for~} u\ll 1 \,.
  \end{eqnarray}
  \label{cutoffextra}
  \end{mathletters}
\par
\begin{figure}  
\epsfbox{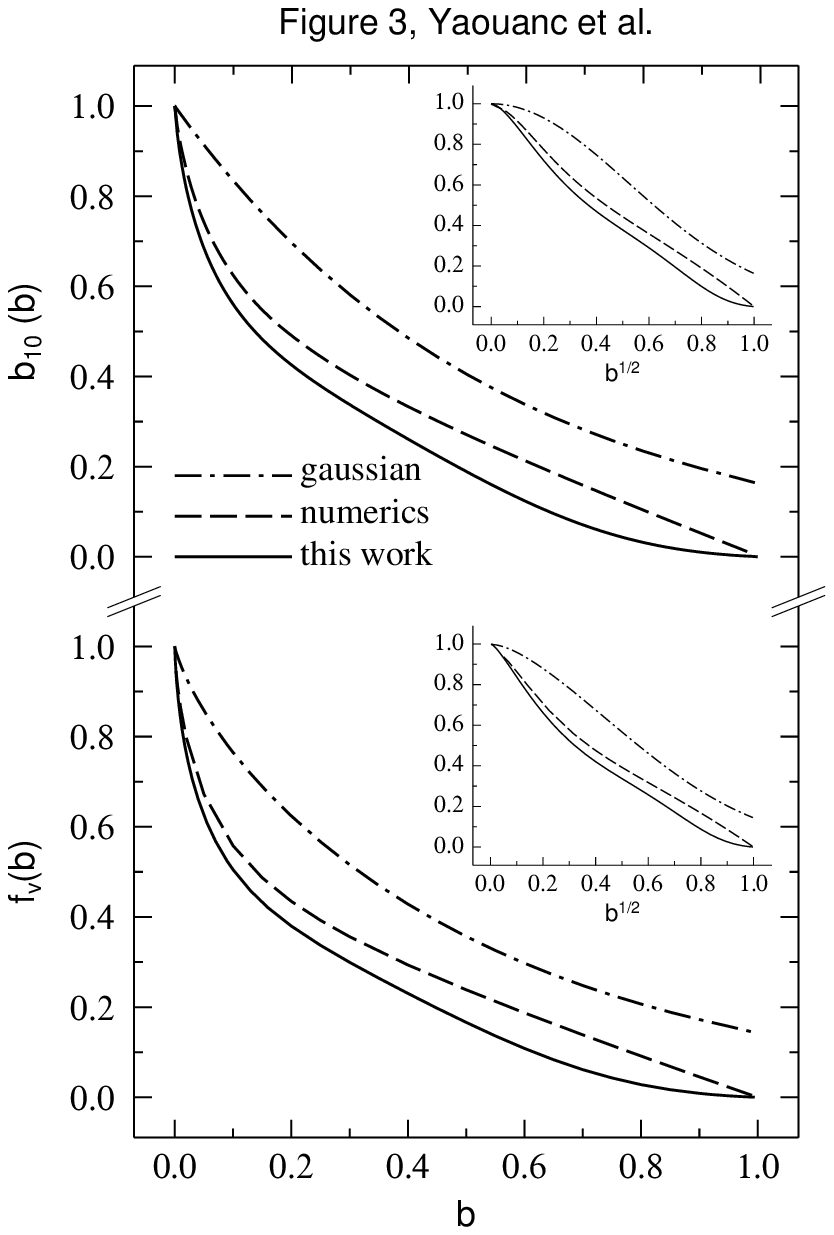}
\caption[]{The universal functions $b_{10}(b)$ (11a) (the largest
reduced form factor, top) and $f_{\rm v}(b)$ (14) (the reduced
variance, bottom) calculated in three ways: From this work (solid
lines), from the Gaussian cutoff (dash-dotted lines), and from the
exact Ginzburg-Landau solution (dashed lines). The inserts plot these
functions versus $\sqrt b$ to stretch the cusp-like $b$ dependence
of the correct cutoff at low reduced inductions $b=0$.
Note the strong deviation of the previously used Gaussian from the
correct cutoff.}
\label{neutronuniversal}
\end{figure}

\noindent For the computation of $B_z({\bf G})$ we need to specify
the geometry of the vortex lattice. As shown by Kogan \cite{Kogan81},
for $B\gg B_{c1}$, the angle characterizing this lattice (see
Fig.\ \ref{lattice}) depends only on the penetration-length ratio :
\begin{eqnarray} 
\tan \alpha & = & {\sqrt 3} (\lambda_x / \lambda_y).
\label{energy5}
\end{eqnarray}
Using Kogan's formula (\ref{energy5}), the form
factor factorizes, $B_z({G_{pq}})$ = $B_0 \cdot b_{pq}(b)$,
where
 \begin{eqnarray}  
B_0 & = & {1 \over \pi^2} \left ( { 3 \over 64} \right)^{1/2}
{\Phi_0 \over \lambda_x \lambda_y}
 \label{extra1}
 \end{eqnarray}
and $b_{pq}(b)$ is a universal function,
  \begin{mathletters} 
  \label{extra2}
  \begin{eqnarray}
b_{pq}(b) & = & {\left(1-b^4 \right)} {v_{pq}\cdot K_1 \left(v_{pq} \right)
\over {p^2 -pq +q^2}},
 \end{eqnarray}
 \begin{eqnarray}
v_{pq} & = &  { 2 \sqrt{2 \pi} \over 3^{1/4} }
{b^{1/2}} \left[ 1 + b^4 \right]^{1/2} \cr
  & \times &  \left[ 1- 2 b \left(1-b^2 \right)^2 \right]^{1/2}
 {\left( p^2 -pq +q^2 \right)^{1/2}}.
 \end{eqnarray}
 \end{mathletters}
In Fig.~\ref{neutronuniversal} we present $b_{10}(b)$ computed from
the variational solution (11), the Gaussian cutoff
(Ref.\ \onlinecite{Brandt72}) and the numerical solution of the GL
equations \cite{Brandtfutur}. Remarkably, the comparison between the
variational and the numerical solutions shows that for $b\le 0.05$
the first three Fourier coefficients $B_z(G)$ deviate by $<10$\%
and for $b\le 0.01$ by $ < 4$\%; even for $b$ = 0.2 (0.3) the
$B_z(G_{10})$ (4) with (5) falls below the exact value by
only 14\% (18\%), and even for small $\kappa =5$ this
Clem-Hao approximation is reasonable.\par

We shall not analyse the SANS data of UPt$_3$ \cite{Kleiman} with
Eq. (\ref{cutoff6}) because the conventional GL theory discussed here
does not describe the phase diagram of this compound. We argue that
the effect of the vortex cores in UPt$_3$ is stronger
than suggested by Joynt \cite{Joyntupt3}.
\par
We now consider the field distribution (probability) of the vortex
lattice which is measured by $\mu$SR \cite{Karlsson} and can be
computed from the Fourier coefficients, Eq. (\ref{cutoff6}). Its
variance is $\Delta_{\text{v}}^2$ =
 $\langle B_z^2 \rangle - \langle B_z \rangle^2$, where
 $\langle \dots \rangle$ means the spatial average. One has
 \begin{eqnarray}   
\Delta_{\text{v}}^2 & = & \sum_{{\bf G} \neq 0} |B_z({\bf G})|^2.
 \label{muon5}
 \end{eqnarray}
$\Delta_{\text{v}}$  separates into two factors,
$\Delta_{\text{v}}$  =  $\Delta_0 \cdot f_{\text{v}} (b)$
where
 \begin{eqnarray}  
\Delta_0 & = &0.06092 {\Phi_0 \over \lambda_x \lambda_y}
 \label{muon7}
 \end{eqnarray}
is the London limit ($\xi_x,\, \xi_y \to 0$) \cite{Barford}
and $f_{\text{v}}(b)$ is a universal function which accounts
for the core size,
  \begin{eqnarray}  
f_{\text{v}}^2(b)  & = & {0.12968 }
\sum_{(p,q) \neq (0,0)} b_{pq}^2 \,,
 \label{muon8}
 \end{eqnarray}
{\it cf.\/}  Fig.~\ref{neutronuniversal}.
The functions $b_{10}$ and $f_{\text v}$ are very similar since
in the sum (\ref{muon8}) the six $b_{10}$ equivalent terms dominate.

Quite unexpectedly, the functions $b_{10}$ and $f_{\text v}$
are {\it strongly field dependent even at low reduced fields $b$},
where the London model predicts constant $b_{pq} = 1$ and
$f_{\text v} = 1$. One has approximately
$1- b_{pq}(b) \propto 1- f_{\text v} \propto b^{1/2}$,
{\it cf.\/}  Fig.~\ref{neutronuniversal}. This finding is confirmed
by the exact numerical solution of the GL theory \cite{Brandtfutur},
depicted as dashed lines in Fig.~\ref{neutronuniversal}. This strong
$b$ dependence originates from the limit (\ref{cutoffextra}c)
with $u^2 = \xi_v^2 G^2 = (8\pi / \sqrt 3) b (G/G_{10})^2$, which
means that the cutoff factor $uK_1(u)$ is considerably less than
unity except at very small $b \ll \sqrt 3 /(8\pi)= 1/14.5 $
even for $G=G_{10}$.

We are aware of only one investigation on a single crystal of the
field dependence of the vortex lattice field distribution
\cite{Sonier}. From this $\mu$SR study of
YBa$_2$Cu$_3$O$_{6.95}$ and the value $B_{c2}$ = 90 (10) T
\cite{Riseman} we estimate $\Delta_{\text{v}}$ $\approx$ 5.04 mT
and 5.73 mT at $B_{\rm ext} =0.5$ T and 1.5 T, respectively.
This leads to a ratio $\cal R_{\text {exp}}$ $\equiv$
 $\Delta_{\text{v}} (1.5\ {\text {T}})/\Delta_{\text{v}}
 (0.5\ {\text {T}})$ = 0.88 while our computation
(see Fig.~\ref{neutronuniversal}) predicts
 $\cal R_{\text {GL}}$ = 0.90. Therefore the conventional GL theory
provides a simple and natural explanation of the observed field
dependence of the observed field distribution in
YBa$_2$Cu$_3$O$_{6.95}$.

In conclusion, we have shown that the effect of the finite core size
on the Fourier components of the magnetic field in a conventional
superconductor with large $\kappa$ is strong, even at
low fields $B_{c1} < B \ll B_{c2}$, since the cutoff factor in Eqs.
(\ref{cutoff4}) and (\ref{cutoff6}) is $uK_1(u) < 1$.
This cutoff effect provides a natural explanation for
recently published neutron and $\mu$SR data without need to resort
to unconventional theories.
\par
\vspace{-.5  cm} 

\end{document}